# Asymmetric $d$-wave superconducting topological insulator in proximity with a magnetic order

M. Khezerlou[a,b], and H. Goudarzi[a]

[a]*Department of Physics, Faculty of Science, Urmia University, P.O.Box: 165, Urmia, Iran*
[b]*National Elites foundation, Iran*

**Abstract**

In the framework of the Dirac-Bogoliubov-de Gennes formalism, we investigate the transport properties in the surface of a 3-dimensional topological insulator-based hybrid structure, where the ferromagnetic and superconducting orders are simultaneously induced to the surface states via the proximity effect. The superconductor gap is taken to be spin-singlet $d$-wave symmetry. The asymmetric role of this gap respect to the electron-hole exchange, in one hand, affects the topological insulator superconducting binding excitations and, on the other hand, gives rise to forming distinct Majorana bound states at the ferromagnet/superconductor interface. We propose a topological insulator N/F/FS junction and proceed to clarify the role of $d$-wave asymmetry pairing in the resulting subgap and overgap tunneling conductance. The perpendicular component of magnetizations in F and FS regions can be at the parallel and antiparallel configurations leading to capture the experimentally important magnetoresistance (MR) of junction. It is found that the zero-bias conductance is strongly sensitive to the magnitude of magnetization in FS region $m_{zfs}$ and orbital rotated angle $\alpha$ of superconductor gap. The negative MR only occurs in zero orbital rotated angle. This result can pave the way to distinguish the unconventional superconducting state in the relating topological insulator hybrid structures.

## 1 INTRODUCTION

The surface states of a three-dimensional topological insulator (3DTI) is considered an intriguing aspect of topological phase of matter. The charge carriers, which are protected by time-reversal symmetry obey from the spin-polarized massless Dirac fermions [1, 2, 3]. Unlike exotic two-dimensional monolayer atomic structures as graphene or molybdenum disulfide, there is no spin and valley degeneracies in single Dirac cone of 3DTI. Among the peculiar properties of topologically conserved surface states one can be addressed: i) theoretically proposed by Fu and Kane [4], and experimentally observed [5, 6] $p$-wave-like superconducting correlations emerged via proximity coupling 3DTI to a conventional $s$-type superconductor ii) the emergence of chiral Majorana mode [7] at the ferromagnet/superconductor (F/S) interface, which is of experimentally importance to detect the Majorana fermions [8, 9]. However, chiral Majorana mode, which corresponds to the zero-energy bound state has a significant impact on the low-energy electron-hole excitations, leading to modifying Andreev reflection (AR) at the F/S interface [10] and thermal transport [11, 12]. On the other hand, anisotropic $d$-wave asymmetry pairing due to including nodal points in its superconducting gap can give rise to potentially forming the zero-energy Andreev bound state and, of course, zero-bias conductance [13, 14]. Actually, angular-resolved AR process and resulting energy bound states at the F/S interface can be influenced by the orbital rotation angle $0 \leq \alpha \leq \pi/4$ of $d$-wave pair potential. However, $Bi$-based cuprate $Bi_{2212}$ is found to be a candidate for anisotropic spin-singlet $d$-wave superconducting gap [15, 16, 17, 18, 19].

Very recently, the coexistence of proximity-induced a conventional superconducting pair potential and a ferromagnetic order at the same time in the surface of 3DTI has been theoretically investigated [20, 21, 22]. The superconducting topological insulator quasiparticle excitations are found to present a new renormalized effective gap by means of a magnetic order. The authors have used spin-singlet $s$-wave order parameter. Due to the spin symmetry of cooper pair, it makes the components to be even in momentum, when the spatial coordinates of two electrons are exchanged. Hence, the exchange of time coordinates has to result in even in frequency. In singlet $d$-wave pairing, the asymmetric pair potential, which is provided by the rotation angle $\alpha$ can be odd in momentum. Therefore, it may be odd under the exchange of time coordinates [23, 24]. Specifically, for maximum orbital rotation angle $\alpha = \pi/4$, the pair potential changes its sign under inversion of angle of incidence $\Delta_d(\pi - \theta_s) = -\Delta_d(\theta_s)$. This causes to create the nodal points in gap. Regarding these aspects, in this paper, we proceed to demonstrate the ferromagnetic order contribution to the asymmetric superconducting effective gap in 3DTI. In conventional superconductor-ferromagnet hybrid (without topological insulator), the interplay between exchange field **M** posed by a magnetic order and superconducting gap gives rise to strongly limiting the magnitude of $|\mathbf{M}|$, so-called Clogston-Chandrasekhar limitation [25, 26]. While, in 3DTI similar hybrid systems, the out-of-plan component of magnetic order $m_z$ opens a gap at Dirac point (no inducing any finite center of mass momentum to the Cooper pair), and odd-frequency triplet component of spin-singlet pair potential creates a gap at Fermi level. The Fermi level is tuned by chemical potential $\mu_{fs}$, that is much larger than superconducting gap. Hence, as an important point, there is no limitation for magnitude of magnetization, and it can take a value up to chemical potential $m_{zfs} \leq \mu_{FS}$ in 3DTI hybrid structures.

Regarding the inversion symmetry breaking at the normal/superconductor interface, the above mentioned superconducting gap may significantly affect the electron-hole conversion, featuring AR process at the interface. This can obviously provide strong changes in charge and spin-polarized transport of ferromagnet/ferromagnetic-superconductor (F/FS) junction. However, the search for transport properties of different hybrid structures including Majorana fermions has led to publish impressive number of guiding theoretical studies for experimental measurements [7, 27, 28, 29, 30, 31]. Particularly, as an interesting feature of topological insulator F/FS interface, we pay attention to the formation of Majorana mode energy with dependency on the induced magnetization. We present, in section 2, the explicit signature of magnetic order in low-energy effective 3DTI $d$-wave superconductor Hamiltonian. The electron(hole) quasiparticle dispersion energy is analytically calculated, which seems to exhibit qualitatively distinct behavior in hole excitations ($|k_{fs}| < K_F$) by varying the magnitudes of magnetization and orbital rotated angle. By considering the magnetization is ever less than chemical potential in FS region, the superconducting wavevector and corresponding eigenstates are derived analytically. Section 3 is devoted to unveil the above key point of FS energy excitation, Majorana mode energy, Andreev process and resulting tunneling conductance and respective discussions. In the last section, the main characteristics of proposed structure are summarized.

## 2 THEORETICAL FORMALISM

### 2.1 3DTI superconducting effective gap

We employ the relativistic generalization of Bogolioubov-de Gennes approach to obtain the electron-hole quasiparticle excitations in the surface state of the 3DTI. These excitations are influenced by magnetic order from a ferromagnet with magnetization **M** on top of the surface of topological insulator. This effect is considered by the perpendicular component of magnetization ($m_{zfs}$) contribution to the Dirac-Bogoliubov-de Gennes (DBdG) Hamiltonian. For finite magnetization, the odd-frequently triplet components, which occur in topological insulator with proximity in a singlet pairing symmetry can become dominant, resulting in a noticeable superconductor subgap structure with emerging low-energy peaks (see, for example Ref. [20, 21, 22]). In Nambu (particle-hole) and spin space, with basis $\psi_{FS}^{e(h)} = \left[\psi_\uparrow(k), \psi_\downarrow(k), \psi_\uparrow^\dagger(-k), \psi_\downarrow^\dagger(-k)\right]^T$, the interplay between ferromagnetic and $d$-wave symmetry superconductor orders in the surface state of the 3DTI is described by the equation

$$\begin{pmatrix} \hat{h}_F^{TI}(k) & \Delta_d(k) \\ -\Delta_d^*(-k) & -\hat{h}_F^{*TI}(-k) \end{pmatrix} \psi_{FS}^{e(h)} = \varepsilon_{FS}^{e(h)} \psi_{FS}^{e(h)}, \tag{1}$$

where $\hat{h}_F^{TI}(k)$ denotes the two-dimensional Dirac Hamiltonian of the 3DTI under the influence of a magnetization **M**

$$\hat{h}_F^{TI}(k) = \hbar v_F (\hat{\boldsymbol{\sigma}} \cdot \mathbf{k}) - \mu_{FS} \hat{\sigma}_0 + \mathbf{M} \cdot \hat{\boldsymbol{\sigma}}.$$

Here, $v_F$ indicates the surface Fermi velocity, and $\mu_{FS}$ is the chemical potential and the ferromagnetic contribution corresponds to an exchange field $\mathbf{M} \equiv (m_x, m_y, m_z)$. Pauli matrices $\hat{\sigma}_0$ and $\hat{\boldsymbol{\sigma}}$ act on spin space. It is worth to point out that the Fermi energy of $Bi$-based 3DTI can be estimated as $\mu_{FS} \geq 50\ meV$. A comparable magnetization can be obtained by proximity from a ferromagnetic insulator. Recently, $EuS$ was experimentally deposited on top of the topological insulator $Bi_2Se_3$ [32], and ferromagnetic order induction to the 3DTI estimated to be around $60 - 400\ meV/nm^2$ per applied Tesla. $\Delta_d(k)$ is superconducting order parameter, which depends on both the orbital and spin-symmetry of the Cooper pair. The gap matrix for spin-singlet $d$-wave asymmetry can be given as

$$\Delta_d^{\pm}(k) = \Delta_0 i \hat{\sigma}_y \cos(2\theta_{fs} \mp 2\alpha) e^{i\varphi}, \qquad (2)$$

where $\Delta_0$ is the uniform amplitude of the superconducting gap, and $\varphi$ is the phase of superconducting order parameter. The $+(-)$ sign is denoted for the case of quasi-electron(quasi-hole), $\theta_{fs}$ is the angle of incidence in superconductor region, and $\alpha$ indicates orbital orientation angular. By diagonalizing the Eq. (1), we arrive at an energy-momentum quartic equation. We suppose the component of magnetization vector along $x$ and $y$-directions to be zero. In this work, we are interesting to induce perpendicular to the surface component of magnetization at the two parallel and antiparallel configurations respective to a ferromagnetic topological insulator, which will be considered in a F/FS junction in the next section.

The dispersion relation resulted from Eq. (1) for electron-hole excitations is found to be of the form:

$$\varepsilon_{FS}^{e(h)} = \zeta \sqrt{\left(-\tau \mu_{FS} + \sqrt{m_{zfs}^2 + v_F^2 |k_{fs}|^2 + \left|\Delta_d^{+(-)}\right|^2 (\frac{m_{zfs}}{\mu_{FS}})^2}\right)^2 + \left|\Delta_d^{+(-)}\right|^2 \left(1 - (\frac{m_{zfs}}{\mu_{FS}})^2\right)}, \qquad (3)$$

where the parameter $\zeta = \pm$ denotes the electron-like and hole-like excitations, while $\tau = \pm$ distinguishes the conduction and valence bands. Of course, equation (3) is clearly reduced to the standard eigenvalues for superconductor topological insulator in the absence of magnetic order $m_{zfs} = 0$, (see, Ref. [27]). The above energy excitation relation is enough complicated. As regards, it is deduced that the effective superconductor subgap is renormalized by magnetization with a factor $\eta = \sqrt{1 - (m_{zfs}/\mu_{FS})^2}$. Indeed, both $m_{zfs}$ and $|\Delta_d|$ cause to dependently open a gap in an otherwise gapless Dirac spectrum of the 3DTI. Also, effective subgap is more or less suppressed for $m_{zfs} \cong \mu_{FS}$ due to the term $\eta \Delta_d$ becomes zero. To satisfy mean field condition that $|\Delta_d|$ must be much smaller than $\mu_{FS}$, then the exact form of superconducting wavevector of charge carriers can be acquired from the eigenstates $k_{fs} = \sqrt{\mu_{FS}^2 - m_{zfs}^2}$. The Hamiltonian of Eq. (1) can be solved to obtain the electron(hole) eigenstates for FS topological insulator. The corresponding wavefunctions including a contribution of both electron-like and hole-like quasiparticles may be found by cumbersome analytical calculations as:

$$\psi_{FS}^e = \begin{pmatrix} e^{i\beta_1} \\ e^{i\beta_1} e^{i\theta_{fs}} \\ -e^{i\theta_{fs}} e^{-i\gamma^e} e^{-i\varphi} \\ e^{-i\gamma^e} e^{-i\varphi} \end{pmatrix} e^{i(k_{fs}^x x + k_{fs}^y y)}, \quad \psi_{FS}^h = \begin{pmatrix} 1 \\ -e^{-i\theta_{fs}} \\ e^{i\beta_2} e^{-i\theta_{fs}} e^{-i\gamma^h} e^{-i\varphi} \\ e^{i\beta_2} e^{-i\gamma^h} e^{-i\varphi} \end{pmatrix} e^{i(-k_{fs}^x x + k_{fs}^y y)} \quad (4)$$

where we define

$$\cos \beta_{1(2)} = \frac{\varepsilon_{FS}^{e(h)}}{\eta \left|\Delta_d^{+(-)}\right|}, \quad e^{i\gamma^{e(h)}} = \frac{\Delta_d^{+(-)}(k)}{\left|\Delta_d^{+(-)}(k)\right|}.$$

It is worth noting the solution is allowed as long as the Zeeman field may be lower than chemical potential, $m_{zfs} < \mu_{FS}$.

## 2.2 N/F/$d$-wave FS junction

Now, we investigate the transport properties of a hybrid N/F/$d$-wave FS junction on the surface ($x - y$ plane) of a 3D topological insulator. The geometry of system is sketched in Fig. 1. The normal metal occupies the region $x < 0$, while the ferromagnetic region with magnetization $m_{zf}$ along $z$-direction extends from $0 \leq x \leq L$, and finally the ferromagnetic $d$-wave superconducting region occupies $x > L$. The superconducting order parameter vanishes identically in N and F regions, and we can neglect its spatial variation in FS region close to the interface. The magnetization vectors of F and FS sections are taken to be at the parallel or antiparallel configurations. In the scattering process follows from the Blonder-Tinkham-Klapwijk (BTK) formula [33], we find the reflection amplitudes from the boundary conditions at the interfaces. It is substantial to determine the allowed values of chemical potentials in three regions. We can set the chemical potential to be zero in F region. The angle of electron (or hole) transmitted to the FS region may be accordingly obtained from the conservation of transverse wavevector under quasiparticle scattering:

$$\theta_{fs} = \arcsin\left(\frac{\mu_N \sin\theta}{\sqrt{\mu_{FS}^2 - m_{zfs}^2}}\right), \tag{5}$$

where $\mu_N$ and $\theta$ are the chemical potential and electron incident angle of N region, respectively. Note that, the electron(hole) angle of incidence in all regions may be adopted in the range $[0, \pi/2]$ around the normal axis to the interface. Regarding the Eq. (5), $\theta_{fs}$ becomes meaningful under conditions that the chemical potential of FS region is greater than magnetic order $m_{zfs} < \mu_{FS}$.

By introducing the transmitted and reflected electron-hole wavefunctions inside the F region

$$\psi_F^{e+} = \left[1, \lambda e^{i\theta_f}, 0, 0\right]^T, \quad \psi_F^{e-} = \left[1, -\lambda e^{-i\theta_f}, 0, 0\right]^T,$$

$$\psi_F^{h+} = \left[0, 0, 1, \lambda e^{i\theta_f}\right]^T, \quad \psi_F^{h-} = \left[0, 0, 1, -\lambda e^{-i\theta_f}\right]^T. \tag{6}$$

we are able to exactly emphasis Andreev process at the F/FS interface. Here, $\lambda = \sqrt{\frac{\mu_F - m_{zf}}{\mu_F + m_{zf}}}$. The probability amplitude of reflections are calculated from the continuity of the wavefunctions at two interfaces. The total wave function in FS region is defined as $\Psi = t^e \psi_{FS}^e + t^h \psi_{FS}^h$. Finally, we find the following analytical expressions for the reflection coefficients

$$\mathcal{R}\left[\varepsilon(m_{zfs}), \theta\right] = \left[t^e e^{i\beta_1}(2M_1 - 1) + t^h e^{-i\beta_2}(2M_2 - 1)\right]\left(i\sin\left(k_f^{xe}L\right)\right) + \left[t^e e^{i\beta_1} + t^h e^{-i\beta_2}\right]\cos\left(k_f^{xe}L\right) - 1, \tag{7}$$

$$\mathcal{R}_A\left[\varepsilon(m_{zfs}), \theta\right] = \left[t^e e^{i\theta_{fs}}e^{-i\gamma^e}(2M_2 - 1) - t^h e^{-i\theta_{fs}}e^{-i\gamma^h}(2M_1 - 1)\right]\left(i\sin\left(k_f^{xe}L\right)\right) - \left[t^e e^{i\theta_{fs}}e^{-i\gamma^e} - t^h e^{-i\theta_{fs}}e^{-i\gamma^h}\right]\cos(k_f^{xe}L), \tag{8}$$

where the parameters read as

$$t^e = \frac{2\Gamma_2 \cos\theta}{\Gamma_2 \Lambda_1 e^{i\beta_1} - \Gamma_1 \Lambda_2 e^{-i\beta_2}}, \quad t^h = \frac{-2\Gamma_1 \cos\theta}{\Gamma_2 \Lambda_1 e^{i\beta_1} - \Gamma_1 \Lambda_2 e^{-i\beta_2}},$$

$$\Gamma_{1(2)} = e^{(-)i\theta_{fs}}\left[(-)N_1\left(M_{2(1)} - 1\right)e^{ik_f^{xe}L} - (+)N_2 M_{2(1)}e^{-ik_f^{xe}L}\right],$$

$$\Lambda_{1(2)} = N_1\left[1 - M_{1(2)}\right]e^{-ik_f^{xe}L} + N_2 M_{1(2)}e^{ik_f^{xe}L},$$

$$N_{1(2)} = \left((-)\lambda e^{(-)i\theta_f} + e^{-i\theta}\right), \quad M_{1(2)} = \frac{\lambda e^{i\theta_f} - (+)e^{(-)i\theta_{fs}}}{2\lambda \cos\theta_f}.$$

The reflection amplitudes measurements under the BTK formalism enables us to capture the tunneling conductance through the junction

$$G(eV) = G_0 \int_0^{\theta_c} d\theta \cos\theta \left(1 + |\mathcal{R}_A[\varepsilon(m_{zfs}), \theta]|^2 - |\mathcal{R}[\varepsilon(m_{zfs}), \theta]|^2\right), \qquad (9)$$

where the critical angle of incidence $\theta_c$ can be determined depending on the doping of F region. The quantity $G_0 \approx N(E_F)we^2/\pi\hbar^2 v_F$ is a normalization factor corresponding to the ballistic conductance of normal metallic junction and $N(E_F) \approx E_F/2\pi(\hbar v_F^2)$ is density of state with $w$ being width of the junction.

## 3 RESULTS AND DISCUSSION

### 3.1 Energy excitation and Majorana mode

In this section, we proceed to investigate in detail the transport properties in normal/ferromagnetic/ferromagnetic superconductor junction on top of a 3DTI, where the superconducting order is spin-singlet $d$-wave symmetry. Due to appearance of nodal points in $d$-wave gap, the superconducting electron-hole quasiparticle excitation around the Dirac points of 3DTI may affect the transport of charge carriers for excitation bellow the effective gap, originated from the Andreev process. In Fig. (2), we demonstrate this excitation when the magnitude of magnetization in FS region $m_{zfs}$ and orbital rotated angle $\alpha$ of superconducting gap can be controlled. In one hand, increasing magnetization causes to disappear the hole excitation, as shown in previous work [22] and, on the other hand, the rotated orbital angle gives rise to rapidly close the superconducting effective gap in Fermi level when $\alpha$ takes a value from zero to $\pi/4$. We set $\alpha = 0$ (solid line), $\alpha = \pi/6$ (dashed line) and $\alpha = \pi/4$ (dashed-dot line), and $m_{zfs} = 0.1\mu_N$ (black) and $m_{zfs} = 0.8\mu_N$ (violet). The energy curves show that Fermi wavevector $K_F$ is displaced toward the lower amounts, $|k| < K_F$ with the increase of $m_{zfs}$. For wavevectors $|k| < K_F$, the dispersion energy curve almost fades out for the higher values of $m_{zfs} \approx \mu_N$. As a result, we expect that the Andreev reflection is suppressed despite superconducting effective gap exists. One of the verified phenomena, which reveals the importance of topological insulator ferromagnet/superconductor interface is the formation of Majorana mode energy. This has a strong relationship with Andreev reflection, and can be achieved when perfect AR occurs. The explicit expression of Majorana energy states is easily given by:

$$\widetilde{\varepsilon}(\theta) = \eta \left( \frac{1 - \Omega^2}{\frac{1}{|\Delta_d^+|^2} + \frac{1}{|\Delta_d^-|^2} - \frac{2\Omega}{|\Delta_d^+||\Delta_d^-|}} \right)^{1/2}, \qquad (10)$$

where

$$\Omega = \cos\left(-i \ln(\frac{\tau_1}{\tau_2})\right),$$

$$\tau_{1(2)} = \Gamma_{1(2)} \left[4iM_{2(1)}\cos\theta \sin(k_f^{xe}L) + 2\cos\theta e^{-ik_f^{xe}L} - \Lambda_{2(1)}\right].$$

We plot $\widetilde{\varepsilon}(\theta)$ as a function of the incident angle for several choices of $m_{zfs}$. It is found that we are able to change the sign of Majorana states by the direction of magnetization $m_{zf}$. This feature of Andreev states provides the chirality of Majorana states. However, as shown in Fig. 3, the Majorana energy changes its sign twice and, in addition to zero incidence, tends to zero for a non-zero incidence of electrons to the interface owing to the electron (hole) incidence angle dependency of $d$-wave pair potential. The position of this new zero-modes can be controlled by the superconductor gap rotated angle and magnetization of FS region. In Fig. 3, we set $\mu_{FS} = \mu_N$ and $\alpha = \pi/4$. The slope of chiral Majorana mode curve near the normal incidence $\theta = 0$ presents smooth with the increase of $m_{zf}$. Consequently, appearance of zero-bias conductance peak may be originated from the flatted bound states near $\theta = 0$. The slope of energy curves around $\theta = 0$ is independent on $m_{zfs}$, while the energy contribution around non-zero angles is redistributed with the increase of $m_{zfs}$. Note that, due to given distribution of Majorana modes the resulting subgap conductance may present a distinctly different behavior.

## 3.2 Conductance

Proposed junction, sketched in Fig. 1, is an ideal setup to measure the conductance spectroscopy like the ones recently performed in Refs. [7, 22, 27]. The conductance difference between parallel and antiparallel configurations of magnetizations in F and FS regions gives rise to capture an experimentally important quantity of magnetoresistance (MR). By using Eq. (9), we calculate numerically the normalized conductance $G/G_0$, as it is commonly done in experiments. Regarding the fact that effective superconducting subgap, renormalized by coefficient $\eta$, varies with the magnitude of magnetization $m_{zfs}$, we may necessarily calculate normal conductance for biases below and above effective subgap. Actually, the effective gap curves (see, Fig. 2) predicates the normalized bias-energy of junction $\varepsilon(eV)/|\eta\Delta_d|$ to reduce, comparing to its common value $\varepsilon(eV)/|\Delta_d| = 1$ in the absence of magnetization. Hence, it needs to exactly determine the relating parameter $\beta$ appeared in Eqs. (7) and (8). Note that, when we integrate the electron-hole reflections probability with respect to the electron incidence angle to measure the subgap conductance, the effective value of paring potential varies with the electron(hole) angle of incidence (see, Eq. (2)). Therefore, we have four conditions for parameter $\beta$ to self adjustingly determine the precise effective superconducting gap respective to the electron(hole) incident with any angle from 0 to $\pi/2$:

$$\beta_{1(2)} = \begin{cases} \arccos\left(\frac{|\varepsilon|}{|\eta\Delta_d^{+(-)}|}\right) & ; \quad |\varepsilon| \leq \left|\eta\Delta_d^{+(-)}\right|, \\ -i\cosh^{-1}\left(\frac{|\varepsilon|}{|\eta\Delta_d^{+(-)}|}\right) & ; \quad |\varepsilon| > \left|\eta\Delta_d^{+(-)}\right|. \end{cases}$$

In fact, these concepts feature asymmetric nature of $d$-wave pairing symmetry. Figures 4 demonstrate the resulting normal conductance of the system. In Figs. 4(a) and (b), the influence of magnetizations in normal conductance is presented. Importantly, as a consequence of unconventional superconductor pairing order, the zero-bias conductance (ZBC) is expected to appear owing to midgap resonant states. For the case of fixed magnetization of F region $m_{zf} = 0.1\mu_N$, the ZBC is strongly enhanced with the increase of magnetization of FS region. While, growing up magnetization of F region $m_{zf}/\mu_N \in [0.1, 0.9]$ leads to strongly decline the ZBC. The former can be understood by the fact that, by increasing $m_{zfs}$, Majorana mode states tend to zero for a non-zero angle of incidence. The main consequence of this key point is that the proportion of zero-energy angles increases in the angle-resolved ZBC. The latter is demonstrated by the gap grown up in Dirac point of surface states owing to the perpendicular component of magnetization. The charge carriers in F region with a high $m_{zf}$, actually, meet a small superconductor subgap due to Fermi wavevector mismatch, and the overgap tunneling occurs almost. For any magnetization of F and FS regions, the overgap ($\varepsilon > \eta\Delta_d$) normal conductance presents flat and small, as expected. Remarkably, the characteristic parameter of $d$-wave symmetry $\alpha$, which plays a role to separate the proportion of electron and hole incidents to the interface inverting the sign of gap for the hole reflection, ($\theta \to \pi - \theta$) shows a notable effect on the subgap tunneling conductance. In zero orbital rotated angle $\alpha = 0$, we find the ZBC peak to fade out for any applied magnetizations to the F and FS regions, see Fig. 4(a).

For the case of antiparallel configuration, we plot in Figs. 4(b) and (c) the normal conductance curves for positive and negative values of $m_{zf}$ in low and high magnetizations of FS region, respectively. There is no difference in resulting ZBC for parallel and antiparallel configurations for low magnetization of FS region. However, these results guide us to achieve the MR of junction via the expression

$$MR(\%) = \left(1 - \frac{G_{\uparrow\downarrow}}{G_{\uparrow\uparrow}}\right) \times 100.$$

Interestingly, tuning the magnetizations of F and FS regions results in both positive and negative MR peaks, as shown in Fig. 5. The positive peak of MR appears when $m_{zfs}/\mu_N$ and $m_{zf}/\mu_N$ becomes maximum and minimum, respectively. Whereas, the scenario for appearance of negative MR is vice versa only in $\alpha = 0$, where the magnetizations of FS and F regions may be minimum and maximum values, respectively. We find no MR in zero bias, which is in agreement with the subgap conductance spectra.

# 4  Conclusion

In summary, in this work we have proposed N/F/FS junction constructed on the surface of 3DTI, where the proximity-induced superconducting order parameter hosts a spin-singlet $d$-wave symmetry. The signature of asymmetric nature of $d$-wave pair potential under electron-hole exchange in the superconducting excitations, Majorana mode energies and resulting subgap and overgap tunneling conductance have been emphasized. The magnetization-induction to the superconducting surface states has caused to appear novel Majorana energy modes at the F/FS interface. The chirality of Majorana modes has been provided by the sign of perpendicular component of magnetization, $\pm m_{zf}$. The ZBC peak has been found to sensitively depend on the $d$-wave characteristic orbital angle and also strength of magnetization of FS region. Considering the normal conductance difference between parallel and antiparallel configurations of magnetizations, we were able to numerically measure the magnetoresistance of junction. Under the assumption of superconductor rotated angle to be zero, a negative MR peak has been achieved by tuning the magnetizations. Since detecting the $d$-wave symmetry state in structures is of experimental importance, these findings can potentially provide a way to distinguish the order parameter.

**Acknowledgments**

The authors would like to thank Vice-presidency and also National Elites foundation of I.R. of Iran for supporting the present work and post-doctorate course of MK at the Urmia University.

**Figure captions**

**Figure 1** (color online) Proposed setup of the topological insulator-based junction that ferromagnetism and $d$-wave superconductivity is induced on the surface of a 3DTI via the proximity effect.

**Figure 2** (color online) The plot of electron-hole excitation for FS 3DTI. We set $\Delta_0 = 0.5\ eV$ and $\theta_{fs} = 0$. The black(violet) lines denote $m_{zfs} = 0.1(0.8)\mu_N$. We choose $\alpha = 0, \pi/6, \pi/4$ for solid, dashed and dashed-dot curves, respectively.

**Figure 3** (color online) Plot of the Majorana modes, which depend on the angle of incidence $\theta_N$ for several values of magnetization in FS section. The parameters of junction are $m_{zf} = 0.4\mu_N, \mu_{FS}/\mu_N = 1$ and $\alpha = \pi/4$.

**Figure 4(a), (b), (c)** (color online) The angle-resolved tunneling conductance versus normalized bias energy. The resulting conductance in N/F/FS is plotted in (a) $m_{zf} = 0.1\mu_N$ for the cases $m_{zfs} = 0.1\mu_N$ (dashed-dot line), $m_{zfs} = 0.6\ \mu_N$ (dashed line), $m_{zfs} = 0.9\ \mu_N$ (solid line), with black lines representing $\alpha = \pi/4$ and brown lines $\alpha = 0$, (b) $m_{zfs} = 0.1\mu_N$ for the cases $m_{zf} = 0.1\mu_N$ (black line), $m_{zf} = 0.8\mu_N$ (brown line) and $m_{zf} = 0.9\mu_N$ (violet line), (c) $m_{zfs} = 0.85\mu_N$ for the cases $m_{zf} = 0.1\mu_N$ (black line), $m_{zf} = 0.6\mu_N$ (brown line) and $m_{zf} = 0.8\mu_N$ (violet line). In (b) and (c) panels, solid lines representing positive values of $m_{zf}$ and dashed lines negative values of $m_{zf}$. We have set $\alpha = \pi/4$.

**Figure 5** (color online) The magnetoresistance as a function of bias voltage for different values of $m_{zf}$. The positive MR peak is obtained for $m_{zfs} = 0.9\mu_N$ and $\alpha = \pi/4$. The negative MR is achieved for $m_{zfs} = 0.1\mu_N$ only in $\alpha = 0$.

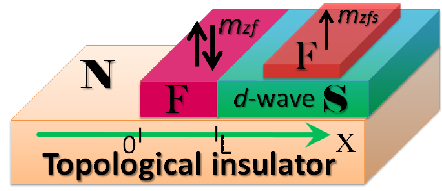

Figure 1:

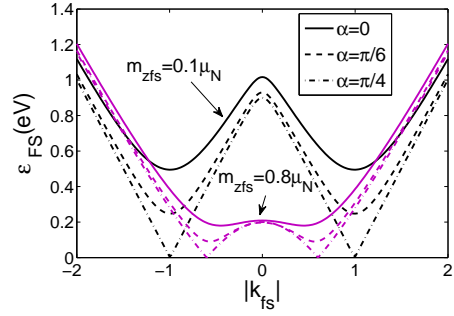

Figure 2:

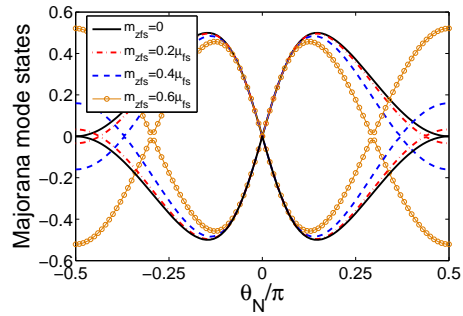

Figure 3:

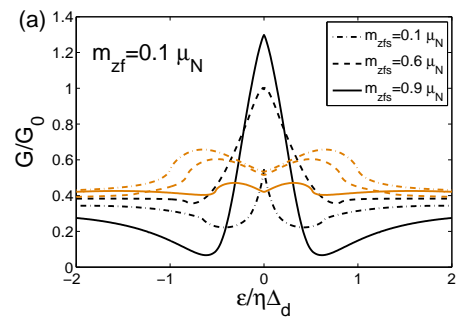

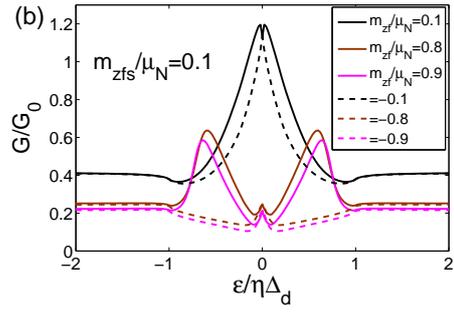

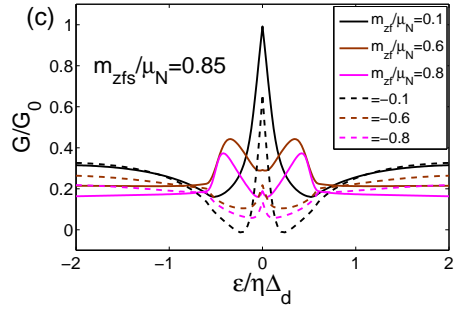

Figure 4: (a),(b),(c)

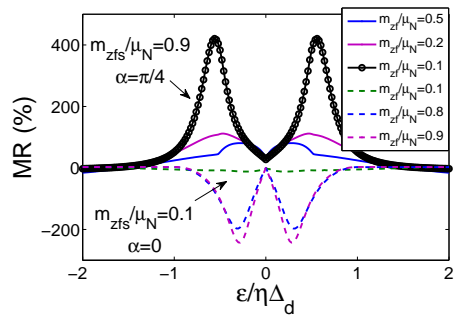

Figure 5: